\documentstyle[12pt,epsf]{article}
\newcommand{\vp}{\varphi}

\newcommand{\lll}{\langle}
\newcommand{\rrr}{\rangle}

\newcommand{\LL}{{\cal L}}
\newcommand{\OO}{{\cal O}}
\newcommand{\HH}{{\cal H}}

\newcommand{\SS}{{\cal S}}

\newcommand{\wt}{\widetilde}

\newcommand{\be}{\begin{equation}}
\newcommand{\ee}{\end{equation}}
\newcommand{\ben}{\begin{eqnarray}\displaystyle}
\newcommand{\een}{\end{eqnarray}}
\newcommand{\refb}[1]{(\ref{#1})}
\newcommand{\p}{\partial}
\newcommand{\sectiono}[1]{\section{#1}\setcounter{equation}{0}}

\begin{document}

{}~ \hfill\vbox{\hbox{hep-th/9911116}\hbox{MRI-PHY/P991133}
}\break

\vskip 3.5cm

\centerline{\large \bf Universality of the Tachyon
Potential}

\vspace*{6.0ex}

\centerline{\large \rm Ashoke Sen
\footnote{E-mail: asen@thwgs.cern.ch, sen@mri.ernet.in}}

\vspace*{1.5ex}

\centerline{\large \it Mehta Research Institute of Mathematics}
 \centerline{\large \it and Mathematical Physics}

\centerline{\large \it  Chhatnag Road, Jhoosi,
Allahabad 211019, INDIA}

\vspace*{4.5ex}

\centerline {\bf Abstract}

Using string field theory,
we argue that the tachyon potential on a D-brane anti-D-brane system in 
type II string theory in arbitrary background has a
universal form, independent of the boundary conformal field theory
describing the brane. This implies that if at the minimum of the
tachyon potential the total energy of the brane antibrane system
vanishes in a particular background, then it vanishes in any other
background. Similar result holds for the tachyon
potential of the non-BPS D-branes of type II string theory, and the
D-branes of bosonic string theory.

\vfill \eject

\baselineskip=18pt

\tableofcontents

\sectiono{Introduction and Summary} \label{s1}

It has been argued on various general grounds that the condensation of the
tachyon living on a configuration of coincident D-brane anti-D-brane
pair
produces a configuration which is indistinguishible from the vacuum where
there are no 
branes\cite{9805019,9805170,9808141,9810188,9812135}. This requires that
the
sum of the tensions of the
brane and the antibrane is exactly cancelled by the (negative) value of
the tachyon potential at the minimum of the potential. There is however
no {\it direct} evidence of this phenomenon, since there is no explicit
knowledge of the tachyon potential, except that it has a maximum
at the origin corresponding to negative mass$^2$ of the tachyon. The
difficulty in studying the tachyon potential can be traced to the fact
that the zero momentum tachyon is far off-shell, and hence is outside the
scope of study of first quantized string theory which deals with only
on-shell S-matrix elements. 

In this paper we shall study some general properties of the tachyon
potential using open string field theory, $-$ a formalism particularly
suited
for the study of off-shell string theory\cite{WITTENSFT,WITTENSUSFT}. In
particular we show that the
tachyon potential on the brane antibrane system is universal, independent
of the particular boundary conformal field theory describing the D-brane, 
except for an overall multiplicative factor which is proportional to the
tension of the brane-antibrane pair before tachyon condensation. Thus for
example, the potential will be the same for flat D-branes,
D-branes wrapped on various cycles of
internal compact manifold, or D-branes in the presence of background
metric and anti-symmetric tensor fields. A similar result holds for the
tachyon potential on a single D-brane of bosonic string theory, or a
single unstable non-BPS D-brane of type II string
theory\cite{9806155,9809111,9812031,9812135}. Although this
does not prove the conjecture that at the minimum of the potential the
tension of the brane antibrane system is exactly cancelled by the tachyon
potential, this shows that if the conjecture is valid for D-brane
anti-D-brane system in one background, then it is also valid for D-brane
anti-D-brane system in any other background.

Let us now be more specific about the analysis and the result of the
paper. Section \ref{s3} of the paper is devoted to the analysis of the
tachyon potential using open string field theory.  As already mentioned,
we shall be interested in a configuration containing a single D-brane in
bosonic string theory, or a D-brane anti-D-brane pair or a single non-BPS
D-brane in type II string theory.  Some of the tangential directions of
the D-brane(s) may be wrapped on some non-trivial cycles of an internal
space. In general such a system of D-branes is described by a non-trivial
boundary conformal field theory. In order to give a uniform treatment of
all systems of this kind, we shall assume that all directions tangential
to the D-brane are compact; this can be easily achieved by compactifying
the non-compact directions tangential to the brane on a torus of large
radii. Thus the resulting configuration can be viewed as a particle like
object in the remaining non-compact directions, which we shall take to be
a Minkowski space\footnote{This restriction is due to a technical reason.
We shall identify the mass of the D-brane as the coefficient of the
${1\over 2}(\dot X^i)^2$ term in the action, and for this purpose we need
some directions in which the space-time is an ordinary Minkowski
space-time.} of dimension $(n+1)$. If we denote the space-like non-compact
directions by $X^i$ ($1\le i\le n$), and the time direction by $X^0$, then
the total world-sheet theory will contain a set of free fields
$X^0,X^1,\ldots X^n$ with Neumann boundary condition on $X^0$ and
Dirichlet boundary condition on $X^1,\ldots X^n$, together with a
non-trivial boundary conformal field theory (BCFT) of central charge
$(25-n)$ describing the dynamics of the coordinates in the compact
direction. The main objective of the paper will be to show that the
tachyon potential is independent of this BCFT.

For simplicity we shall focus our attention on D-branes of bosonic string
theory during most of the paper; so let us explain our results first in
this context. We shall show that tachyon potential has the
form:\footnote{Throughout this paper all masses and energies will be
measured in the closed string metric.}
\be \label{e2.1}
V(T)= M f(T)\, ,
\ee
where $f(T)$ is a universal function of the tachyon field $T$ independent
of the BCFT describing the D-brane, and $M$ is the mass of the D-brane at
$T=0$,
which can depend on the BCFT under consideration.\footnote{In the
convention that we shall choose, the mass of the D-brane is
also independent of the BCFT. However it depends on the open string
coupling constant, whose relation to the closed string coupling constant
may depend on the details of the BCFT.} During this analysis we shall
also arrive at a precise definition of the tachyonic mode(s) and the
tachyon potential.
We choose the additive constant in $V(T)$ such that it vanishes at
$T=0$. Thus
the total energy of the D-brane for a given value of $T$ will be given by
\be \label{e2.2}
M+V(T)= M \{ 1+f(T)\}\, .
\ee
According to the conjecture of \cite{9902105,RECK}, at some extremum
$T_0$ of the
tachyon potential the negative contribution of the tachyon potential
exactly cancels the mass of the D-brane. Thus according to this conjecture
\be \label{efvan}
1+f(T_0)=0\, .
\ee
Although our
analysis does not provide a proof of this relation, the universality of
the function $f(T)$ shows that if the
relation holds for any of the D-branes of the bosonic string theory (say
the D0-brane of the bosonic string theory in 26 dimensional Minkowski
space), then it must hold for all D-branes in all possible
compactifications of bosonic string theory. 

An exactly similar result holds for the brane
antibrane system of type II string theory. In this case $M$ denotes the
total mass of the brane-antibrane
system under consideration. The function $f(T)$ differs from the
corresponding function in the bosonic string theory, but it is again
universal in the sense that it does not depend on the details of the BCFT
describing the brane antibrane system. The conjecture of
ref.\cite{9805019,9805170}
again requires $\{1+f(T)\}$ to vanish at an extremum $T_0$ of
$f(T)$. This time, however, supersymmetry of the background space-time
requires that $T_0$ satisfying eq.\refb{efvan} represents a global minimum
of the potential.

Finally the result also holds for the non-BPS D-brane of type II string
theory, with $M$ now representing the mass of the non-BPS D-brane.

According to the conjecture of \cite{9805019,9805170,9812031,9902105}, at
$T=T_0$
the D-brane of bosonic string theory,
the brane
antibrane
system of type II string theory, or the non-BPS D-brane of type II string
theory, is
indistinguishible from the vacuum where there is no D-brane. 
Since the tachyon is neutral under the `center of 
mass' U(1) gauge
field living on the brane (brane antibrane system), a vev of the tachyon
field does not break this
U(1) gauge symmetry.  On the other hand the vacuum without any D-brane
does not
contain such a U(1) gauge field. This poses a
puzzle\cite{9807138,9810188,9901159}. In section \ref{s4}
we show that the results of section \ref{s3} points to a possible way out
of this puzzle.  
Using the universality of the tachyon potential, and the fact that
$(1+f(T_0))$ vanishes at $T_0$, we argue
that
at $T=T_0$, the effective action involving the center of mass U(1) gauge
field does not contain any term without derivative of the {\it gauge field
strength}. In particular it implies that the standard gauge kinetic term
is absent. We conjecture that the effective action at $T=T_0$ is
altogether independent of the gauge field, so that the gauge
field behaves as an auxiliary field.
This would explain the absence of a dynamical U(1) gauge field at $T=T_0$.
Its equations of motion forces
all states carrying the U(1) charge to disappear from the
spectrum.\footnote{The argument given in this section is an expanded
version of the analysis already presented in \cite{9909062}.}

Finally in section \ref{s5} we discuss generalization of our results to
closed
bosonic string theory. We show that arguments similar to the one given in
section \ref{s3}  can be used to establish the universality of the tachyon
potential in any compactification of the bosonic string theory. However,
since there is no compelling reason to believe that there is a stable
minimum of this potential, the significance of this result is not entirely
clear.

Although our analysis establishes the universality of the tachyon
potential, it does not tell us what this universal function is.
Explicit analysis
of the tachyon potential in open string theory with all Neumann boundary
conditions was carried out in
ref.\cite{BERD}. Some
properties of the tachyon potential on the brane antibrane system have
been analyzed previously in refs.\cite{9403040,9511194,9902181}. Attempts
at
deriving the explicit form of the tachyon potential using open string
field theory have been made earlier in refs.\cite{KOST}. Similar analysis
for closed string tachyons were carried out in
refs.\cite{KOST2,9409015,9412106}. 
Some aspects of the universality of the tachyon potential have been
addressed earlier in ref.\cite{BANKS}.

\sectiono{Tachyon potential from open string field theory on the D-branes}
\label{s3}

We shall use Witten's open string field theory\cite{WITTENSFT,WITTENSUSFT}
to analyse the tachyon
potential, but any other formulation of covariant open string field theory
will
also suffice\cite{9705241}. Although the original version of this theory
was formulated
for open strings in flat space-time with Neumann boundary conditions in
all directions, it can be easily generalized to describe open strings
living on a D-brane. We use the language of \cite{LPP}, as
reformulated in \cite{SE} for describing string field theory in arbitrary
background field. We shall begin our discussion with open strings
living on a
D-brane in bosonic string theory; and later generalise it to
brane-antibrane system or non-BPS D-branes in superstring theories.

As mentioned in the introduction, we compactify all the spatial directions
tangential to the D-brane. Thus we are dealing with the dynamics of a
particle with infinite number of degrees of freedom, described by a (0+1) 
dimensional string field theory.
Since string field theory corresponds to second quantized string theory,
a point in the {\it classical} configuration space of string field theory
corresponds to a
specific {\it quantum}
state of the first quantized theory. As was shown in \cite{WITTENSFT}, in
order to describe a gauge invariant string field theory we must include
the full Hilbert space of states of the first quantized open string
theory
including the $b$ and $c$ ghost fields, subject to the
condition that
the state must carry ghost number 1. Here we are using the
convention that $b$ carries ghost number $-1$, $c$ carries ghost number 1,
and the SL(2,R) invariant vacuum $|0\rrr$ carries ghost number 0.
We shall denote by $\HH$ the subspace of the full Hilbert space
carrying ghost number 1. Let
$|\Phi\rangle$ be an arbitrary state in $\HH$, 
and $\Phi(x)$ be the local field (vertex operator) in
the
conformal field
theory 
which creates this state $|\Phi\rrr$ out of the SL(2,R) invariant vacuum:
\be \label{e3.1a}
|\Phi\rrr=\Phi(0)|0\rrr\, .
\ee
Since we are dealing with open string theory, $\Phi(x)$ lives on the
boundary of the world sheet.
We shall choose the convention that the world-sheet is the
upper half plane, and its boundary is the real axis labelled by $x$.

The open
string field
theory action, which is a map from $\HH$ to the space of real numbers, is
given by
\be \label{e3.1}
\SS =-{1\over g_o^2} \bigg({1\over 2} \lll\Phi|Q_B|\Phi\rrr +{1\over 3}
\lll
f_1\circ\Phi(0)
f_2\circ\Phi(0) f_3\circ\Phi(0)\rrr\bigg)\, .
\ee
Here $g_o$ is a constant denoting the open string coupling constant, 
$Q_B$ is the BRST charge constructed out of the ghost oscillators
and the matter stress tensor, and $\lll~\rrr$ denotes correlation
functions in
the combined matter and ghost conformal field theory. The overall
$-$ sign in front of the action is a reflection of the fact that we
are using Minkowski metric with signature $(-++\ldots +)$. $f_1$, $f_2$
and $f_3$ are three conformal transformations given by, \ben \label{e3.2}
&& f_1(z)=-e^{-i\pi/3}\bigg[\bigg({1-iz\over 1+iz}
\bigg)^{2/3}-1\bigg]\bigg/ \bigg[\bigg({1-iz\over 1+iz}
\bigg)^{2/3}+e^{i\pi/3}\bigg], \nonumber \\
&& f_2(z)=F(f_1(z)), \qquad
f_3(z)=F(f_2(z))\, ,
\een
where $F$ is an SL(2,R) transformation
\be \label{e3.2a}
F(u)=-{1\over 1+u}\, .
\ee
$f_i\circ\Phi(0)$ denotes the conformal transform of $\Phi(0)$ by
$f_i$. Thus for example if $\Phi$ denotes a dimension $h$ primary field,
then $f_i\circ\Phi(0)=(f_i'(0))^h\Phi(f(0))$. For non-primary fields there
will be extra terms involving higher derivatives of $f_i$. The inner
product
appearing in the first term of the action is defined as
\be \label{ebpz}
\lll\Phi|\Psi\rrr = \lll I\circ\Phi(0)\Psi(0)\rrr
\ee
where $I$ denotes the SL(2,R) transformation $I(z)=-(1/z)$.
We shall choose the convention where $\alpha'=1$, and the SL(2,R)
invariant vacuum $|0\rrr$ is normalized as
\be \label{enorm}
\lll 0|c_{-1} c_0 c_1|0\rrr = L\, ,
\ee
$L$ being the (infinite) length of the time interval over which the
action
is evaluated. (For the purpose of normalization we shall pretend
that the time direction is compact with radius $L/2\pi$.) $c_n$ are the
modes of the ghost field $c(z)$ defined through the relation $c(z)=\sum
c_n z^{-n+1}$. In general we normalize the Fock vacuum $|k_0\rrr\equiv 
\exp\big(ik_0X^0(0)\big)|0\rrr$ with $X^0$ momentum
$k_0$ as
\be \label{enorm1}
\lll k_0|c_{-1} c_0 c_1 |k_0'\rrr = 2\pi \delta(k_0+k_0')\, ,
\ee
with the understanding that $\delta(0)$ is defined to be $L/2\pi$.

The equations of
motion of string field theory are obtained by demanding that the variation
of $\SS$ with respect to $|\Phi\rrr$ vanishes. We can get the component
form 
of the equations by decomposing $|\Phi\rrr$ in a complete set of basis
states in $\HH$, and setting to zero the variation of $\SS$ with respect
to each coefficient in this expansion.

The zero momentum tachyonic state of open string theory can be identified
as
\be \label{e3.3}
c_1|0\rrr\, ,
\ee
created by the vertex operator $c(0)$ acting on $|0\rrr$.
It is however clear that due to the cubic coupling in the string field 
theory action \refb{e3.1}, once we switch on tachyon vacuum expectation
value (vev), various other fields must also be switched on in order to
satisfy the string field theory equations of motion. However, not all the
fields need to be switched on.
Suppose we can decompose $\HH$ into two subspaces $\HH_1$ and
$\HH_2$ such that $\SS$ is always quadratic or higher order in the
components
of $|\Phi\rrr$ along the basis vectors of $\HH_2$. If we now take
$|\Phi\rrr$ to
lie solely in $\HH_1$, then all the equations of motion obtained by
varying $\SS$ with respect to the components of $|\Phi\rrr$ along $\HH_2$
are
automatically satisfied. Thus we can obtain a consistent truncation of the
theory by restricting $|\Phi\rrr$ to $\HH_1$ and evaluating $\SS$ for this
$|\Phi\rrr$. A solution of the equations of motion obtained by varying the
truncated action
with respect to comoponents of $|\Phi\rrr$ along $\HH_1$ can automatically
be
regarded as a solution of the equations of motion of the full string field
theory.

We shall now describe such a decomposition of $\HH$. 
We include in $\HH_1$ all states of ghost number 1, obtained from the
SL(2,R) invariant vacuum by the action of the
ghost oscillators $b_n$ and $c_n$, and the Virasoro generators of the
entire matter conformal field theory. In the language of vertex operators
this will amount to including those vertex operators which can be
obtained as products of (derivatives of)  $b(x)$, $c(x)$, and the
matter
stress tensor $T^{(matter)}(x)$. $\HH_2$ will contain all states of
ghost number 1 carrying non-zero $k_0$, and also all states with
$k_0=0$ which are obtained by the action of $b_n$, $c_n$ and the
matter Virasoro generators on primary states of dimension $>0$ of the
matter conformal field theory. Since the BRST operator $Q_B$
is constructed from the ghost oscillators and  matter Virasoro generators,
the kinetic term of the action \refb{e3.1} does not mix a state in
$\HH_1$ with a state in $\HH_2$. A conformal transformation takes
a state in $\HH_1$ ($\HH_2$) to a state in $\HH_1$ ($\HH_2$), and
furthermore, the three point correlation function
of two vertex
operators in $\HH_1$
and a vertex operator in $\HH_2$ vanishes.
Thus
restricting the string field configuration to $\HH_1$ will give a
consistent truncation of the string field theory.

Since the zero momentum tachyon state described by eq.\refb{e3.3} belongs
to
$\HH_1$, we see that switching on this tachyonic mode does not take us
outside the subspace
$\HH_1$. In particular the tachyonic ground state
will correspond to a state $|\Phi_0\rrr$ with no
component along $\HH_2$, and satisfying the equations of motion
derived from the truncated action. Since integrating out all the modes in
$\HH_1$ other than $c_1|0\rrr$ may not lead to a meaningful
approximation,\footnote{Indeed, the true ground state may not have any
component along $c_1|0\rrr$.} we denote by the single symbol $T$ the
set of all the modes of $\HH_1$, and by $\wt \SS(T)$ the truncated string
field
theory
action, with the string field
configuration $|\Phi\rrr$ restricted to $\HH_1$. 
Since $\HH_1$ involves only those states which carry zero $X^0$ momentum,
the inner product as well the three point function appearing in 
eq.\refb{e3.1} will contain a $\delta(0)$ term, representing the
infinite contribution from the time integral of a time independent
lagrangian. Thus the lagrangian $\wt
\LL(T)$ for this configuration can
be identified as the action $\wt \SS(T)$ with this volume factor
$L=2\pi\delta(0)$
removed. Once $\wt \LL$ has been constructed this way, the
tachyonic potential $V(T)$ can be identified with $-\wt \LL(T)$.

Computation of $V(T)$ only involves correlation functions 
involving the ghost fields and the matter energy momentum tensor with
central charge 26. These
correlation functions are completely universal. In particular, they are
insensitive
to all the details of the internal BCFT. As a result, $V(T)$ has a
universal
form for all internal BCFT except for the overall multiplicative
factor $g_o^{-2}$ in front of the action \refb{e3.1}.
Thus the tachyon
potential has the form: 
\be \label{e3.4} 
V(T)={1\over g_o^2} h(T)\, , 
\ee
where $h(T)$ is some universal function independent of the choice of the
internal BCFT. 

We shall now show that at $T=0$ the mass of the D-brane described by the
action
\refb{e3.1} is related to $g_o^{-2}$. To see this
let us consider the kinetic term in \refb{e3.1} involving the mode
$\int dk_0 \phi^i(k_0) c_1 \alpha^i_{-1} |k_0\rrr$. Here $\alpha^i_{n}$
denotes the oscillator
of the free world-sheet scalar field $X^i$, and $|k_0\rrr$ denotes the
state $\exp(ik_0 X^0(0))|0\rrr$. Only the $c_0 L_0^{matter}$ term of the
BRST charge $Q_B$ contributes to the $k_0$ dependent part of the kinetic
term involving this mode, and
the result is given by
\be \label{e3.5}
2\pi \, {1\over 2} (g_o)^{-2}\int dk_0 (k_0)^2 \phi^i(k_0)\phi^i(-k_0)\, ,
\ee
in the $\alpha'=1$ unit. If $\psi^i(t)\equiv \int dk_0 e^{i k_0 t}
\phi^i(k_0)$ denotes the Fourier transform of
$\phi^i(k_0)$, then the above action can be rewritten as
\be \label{e3.6}
{1\over 2} (g_o)^{-2}\int dt \p_t \psi^i \p_t \psi^i\, ,
\ee
where $t$ denotes the time variable conjugate to $k_0$. Up to an overall
normalization factor, 
$\psi^i$ has the interpretation of the location of the D-brane in the
$x^i$
direction. This normalization factor may be determined as follows. Instead
of taking a single D-brane, let us take a pair of identical D-branes,
separated by a distance $b^i$ along the $X^i$ direction. Then each state
in the open string Hilbert space carries a $2\times 2$ Chan Paton factor,
and states with off diagonal Chan Paton factors, representing open strings
stretched between the two branes, are forced to carry an
amount of winding charge $b^i$ along $X^i$. If we now move one of the
branes by an amount $Y^i$ along $X^i$, the change in the (mass)$^2$ of the
open string with Chan Paton factors
$\pmatrix{0 & 1\cr 0 & 0}$ and $\pmatrix{0 & 0\cr 1 & 0}$ should be
given by:
\be \label{ecpone}
{1\over (2\pi)^2} \{(\vec b+\vec Y)^2 -\vec b^2\} = {1\over 2\pi^2} \vec
b\cdot \vec Y + O(\vec Y^2)\, .
\ee
In the above equation we have used the fact that with our choice of units,
the string
tension is equal to $(1/2\pi)$.
On the other hand, since $\psi^i$ denotes the mode which translates the
brane, 
moving {\it one of the branes} along $X^i$ will correspond to
switching on a constant $\psi^i$. This is represented by a string field
background 
\be \label{ecptwo}
\psi^i c_1\alpha^i_{-1}
|0\rrr\otimes \pmatrix{1 & 0\cr 0 & 0}\, .
\ee
We can now explicitly use the string field theory action \refb{e3.1} to
calculate the
change of the (mass)$^2$ of states with Chan Paton factors $\pmatrix{0 &
1\cr 0
& 0}$ and $\pmatrix{0 & 0\cr 1 & 0}$ due to the presence of this
background string field. The result is
\be \label{ecp3}
{1\over \sqrt 2\pi} \vec b\cdot \vec \psi + O(\vec \psi^2)\, .
\ee
Comparing eqs.\refb{ecpone} and \refb{ecp3}
we get
\be \label{ecp4}
\psi^i={Y^i\over \sqrt 2\pi}\, .
\ee
Once we have determined the relative normalization between $\psi^i$ and
$Y^i$, we can return to the system containing a single
brane.\footnote{This
can be done, for example, by moving the other brane infinite distance away
by taking the limit $|\vec b|\to\infty$.} Substituting eq.\refb{ecp4}
into eq.\refb{e3.6}, we get,
\be \label{ecp5}
{1\over 2} (g_o)^{-2} (2\pi^2)^{-1}\int dt \p_t Y^i\p_t Y^i\, .
\ee
This contribution to
the D-brane world-volume action can be
interpreted as due to the kinetic energy associated with
the collective motion of the D-brane in the non-compact transverse
directions. This allows
us to identify the D-brane mass as
\be \label{e3.7}
M=(2\pi^2)^{-1}(g_o)^{-2}\, .
\ee
Thus eq.\refb{e3.4} can be rewritten as
\be \label{e3.8}
V(T)=M f(T)\, .
\ee
where $f(T)\equiv 2\pi^2 h(T)$ is another universal function.
This proves eq.\refb{e2.1} for the tachyon potential on a single bosonic
D-brane.

Let us now consider the case of tachyon condensation on a brane-antibrane
pair in type II string theory. Since the analysis is very similar to
the case discussed above, we shall only point out the essential
differences. Open string field theory with cubic action
has been constructed in \cite{WITTENSUSFT}. The string field contains two
separate components, one from the Neveu-Schwarz (NS) sector and the other
from the Ramond (R) sector; but for the study of tachyon potential we can
set to zero the R sector fields. A generic NS sector string field
configuration
is a state in the Hilbert space $\HH$ of the form $\Phi(0)|0\rrr$, where
$|0\rrr$ denotes the SL(2,R) invariant vacuum, and $\Phi(x)$ is the
product of $e^{-\phi(x)}$ with an arbitrary operator $\OO(x)$ of ghost
number 1, made from products of (derivatives of) $b$, $c$, the bosonic
ghost fields $\beta$, $\gamma$, and matter operators. The
ghost charge is defined such that $b$ and $\beta$ carry ghost number
$-1$ and $c$ and $\gamma$ carry ghost number 1.
$\phi$ denotes the scalar field obtained by `bosonizing' the
$\beta-\gamma$
system\cite{FMS}. In the left hand side of the normalization condition
\refb{enorm} we now need to include an additional factor of
$e^{-2\phi(0)}$ besides the $c_{-1}c_0c_1$ factor. There is a further
subtlety due to the fact that $\HH$ contains four sectors labelled by
the $2\times 2$ Chan Paton (CP) factor. We shall take the identity
matrix $I$ and three Pauli matrices $\sigma_i$ to be the four
independent CP factors. States in the CP sector $I$ and $\sigma_3$ satisfy
the conventional GSO projection rules according to which $|0\rrr$ is even, 
and $e^{-\phi}$, $\beta$, $\gamma$ are odd. States in the CP sector
$\sigma_1$ and $\sigma_2$ satisfy the opposite GSO projection rules
according to which $|0\rrr$ is odd. 
The
tachyon field is complex, but we shall restrict to configurations
with real tachyon background. The zero momentum
tachyon field then corresponds to the state created by the vertex operator 
$c(0)e^{-\phi(0)}\otimes \sigma_1$ on $|0\rrr$.

The string field theory action has a
form very
similar to \refb{e3.1}, with the difference that the cubic interaction
vertex also contains an insertion of the picture changing
operator\cite{FMS} in the correlation function. Since this operator
involves only ghost fields and the super-stress tensor of the matter
fields, it is independent of the choice of BCFT describing the brane
antibrane pair and will not affect our argument. 
As in the case of bosonic open string field theory, we can obtain a
consistent truncation of the string field theory action by restricting
$|\Phi\rrr$ to states for which the corresponding vertex operator
$\Phi(x)$
is built from products of (derivatives of) the ghost fields, and the
matter
super stress tensor. This includes the energy momentum tensor
$T^{(matter)}(x)$ and
the supercurrent $G^{(matter)}(x)$.  Furthermore since $(\sigma_1)^2$ is
the identity matrix $I$, we can restrict ourselves to states with CP
factors $I$ and $\sigma_1$ only, with the usual GSO projection on the
states with CP factor $I$, and opposite GSO projection on the states with 
CP factor $\sigma_1$. The resulting truncated action is again universal,
and in particular
insensitive to the details of the internal BCFT.
This shows that the tachyon potential has the form \refb{e3.4} for some
universal function $h(T)$. (This of course is different from the universal
function which appears in bosonic string theory.) Furthermore the mass of
the D-brane is still given by an equation similar to \refb{e3.7}. Thus
$V(T)$ has the form
given in eq.\refb{e3.8}.

One of the crucial assumptions in our argument is that the BCFT describing
the D-brane anti-D-brane system has a factorized form so that the
conformal field theory describing the open strings is
identical in each of the four CP sectors (except for opposite GSO
projections in sectors $\sigma_1$ and $\sigma_2$). In particular,
$e^{-\phi(0)} c(0)|0\rrr\otimes\sigma_1$ must be an allowed state
in the theory. Formally this can
be achieved if the antibrane is always defined to be the configuration 
obtained from the brane by the operation of
$(-1)^{F_L}$, where $(-1)^{F_L}$ denotes the transformation which changes
the sign of all the R-R and
R-NS sector closed string states. In the language of boundary
states this means that the antibrane is defined to have the same
boundary state as the brane, except that the sign of all the RR
states is reversed. However we should keep in mind that it
is certainly possible to construct brane-antibrane system which does not
fall into this category. 
A simple example would be brane-antibrane pair separated by a
distance $b$ in a direction transverse to the brane. In this
case the states in the
CP sector $\sigma_1$ and $\sigma_2$ are forced to carry non-zero string
winding charge proportional to $b$, and hence the string field
configuration describing a zero momentum tachyon background  
is no longer of the form $c(0)e^{-\phi(0)}|0\rrr\otimes \sigma_1$. Instead
it corresponds to a state built from a non-trivial primary state of the
BCFT. Thus our argument for the universility of the tachyon
potential is no longer valid in this case. A similar situation arises, for
example, if either the brane or the antibrane (but not both) carries a
Wilson line or a magnetic field tangential to its world volume.

A very similar argument can be given for the universality of the tachyon
potential on a non-BPS D-brane of type II string theory. In fact, since
the non-BPS D-brane of type IIB (IIA) string theory can be regarded as the
result of modding out a brane-antibrane pair of type IIA (IIB) string
theory by $(-1)^{F_L}$\cite{9812031}, the universality of the tachyon
potential
on a brane-antibrane system of type II string theory automatically implies
the universality of the tachyon potential on a non-BPS D-brane of type II
string theory.

The analysis of this section indicates that it should be possible to
describe the string field configuration corresponding to $T=T_0$ as a
universal state in
$\HH_1$. This state should represent a solution of the classical equations
of motion of string field theory, and should have the property that when
we analyze small fluctuations of string field around this solution, the
spectrum should not contain any physical states. (This is necessary if
we are to interprete the configuration $T=T_0$ as the vacuum without any 
brane.) We should caution the
reader however that our arguments are quite formal, since {\it a priori}
there is no reason to expect that the $T=T_0$ configuration can be
represented as a normalizable state in $\HH_1$. Nevertheless, formal
solutions of string field theory equations of motion have provided
valuable insight in the past\cite{CUBIC,CUBIC2}. In fact,
ref.\cite{CUBIC2} does contain examples of such formal solutions which
do not have any physical excitations. Finding a (formal) solution
of the string field theory equations of motion
which satisfies eq.\refb{efvan}, and hence represents the vacuum
state, remains an open problem.

We end this section by noting that the result of this section has been
implicitly used in ref.\cite{9810188} in classifying D-branes via
K-theory. Universality of the tachyon potential, together with
eq.\refb{efvan}, shows that a brane and an antibrane can always annihilate
via tachyon condensation as long as their boundary states differ from
each other just by a change of sign of the Ramond-Ramond states. This
requires that they carry the same gauge
bundle, {\it i.e.} that only gauge fields with CP factor $I$ are excited.
Such
brane-antibrane annihilation forms a crucial ingredient in establishing
one
to one correspondence between stable D-branes and elements of the K-group.

\sectiono{Fate of the U(1) gauge field under tachyon condensation}
\label{s4}

In this section we shall use the results of the previous section to
discuss the fate of the U(1) gauge field on the D-brane under tachyon
condensation. The salient points of this analysis were already given in
\cite{9909062}.

Let us begin with the bosonic D-brane. There is a U(1) gauge field living
on the D-brane. The tachyon is neutral under the gauge group; hence our
intuitions from quantum field theory will tell us that the gauge fields
will remain massless even when the tachyon condenses. On the other hand if
$T=T_0$ corresponds to the vacuum without any
D-branes, as has been conjectured,
then clearly there cannot be a U(1) gauge field living on the brane after
tachyon condensation. How do we resolve this apparent contradiction? A
related question is as follows. If we consider a pair of D-branes (not
necessarily of the same kind)
separated by a distance, then there is an open string state with one end
on each brane. If we now let the tachyon on one of the branes condense,
then what happens to this open string state? If the $T=T_0$
configuration
really represents the vacuum, then there cannot be an open string ending
at the original location of the brane after tachyon condensation. 

The resolution that we propose is as follows. We conjecture that at
$T=T_0$ the action of the U(1) gauge
field on the D-brane world volume is independent of the gauge
field. In fact, we conjecture that the action is independent of all
the massless fields living on the D-brane world volume. Thus the gauge
field is no
longer dynamical, but acts as an auxiliary field which forces the
corresponding U(1) current to vanish identically. In particular this means
that open strings with one end on this brane and another end on some other
brane, being charged under the U(1), is no longer a physical state.
Physically this can be explained by saying that since effectively the U(1)
gauge
coupling becomes infinite, any state charged under this U(1) becomes
infinitely massive and hence decouples from the spectrum.\footnote{This
interpretation makes contact with the conjecture of ref.\cite{9901159}
that this U(1) gauge field is confined.}

Although we have no general proof of this statement, we shall now show
that our analysis of the
previous section can be used to lend support to this conjecture. 
For this, let us start with a D-$p$-brane of the bosonic string theory,
and  compactify all directions
tangential to the
brane on a torus $T^p$ of large radii. Let $\vec y$ denote the directions
tangential to the brane, $\{\vp_a(\vec y)\}$ denote an arbitrary time
independent configuration of all
massless
fields living on the brane world-volume, and $T$ denote the tachyonic
mode(s)
discussed in the last section. We denote by $\LL(\{\vp_a(\vec y)\},T)$
the effective lagrangian of the brane obtained by integrating out all
other modes. Note that $T$ correspond to mode(s) carrying zero momentum
along the world-volume direction, whereas the massless fields
$\{\vp_a(\vec y)\}$ are allowed arbitrary dependence on the
world-volume coordinates. All other modes have been integrated out. This
would typically give an effective lagrangian which is non-local on the
D-brane world-volume, but this will not affect our discussion.

\begin{figure}[!ht]
\leavevmode
\begin{center}
\epsfbox{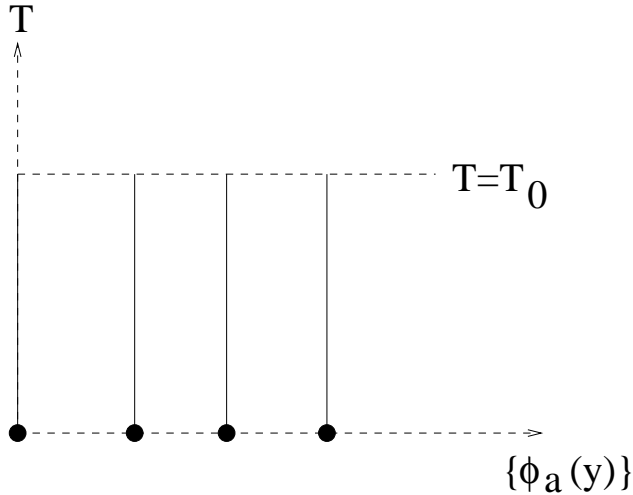}
\caption{This diagram schematically illustrates the choice of coordinate
system in the configuration space. The horizontal axis denotes the set of 
all time independent configurations of massless fields, and the vertical
axis denotes the tachyonic mode(s)
$T$. The black dots on the horizontal axis are the classical solutions of 
the equations of motion involving massless fields only. The vertical
line originating from a black dot 
represents the effect of
switching on the tachyonic mode(s) of the string field theory formulated
around  the BCFT associated with the particular black dot.}
\label{f1} 
\end{center} 
\end{figure}
At this point we need to make some further remark about the choice of the
coordinate $T$ in the configuration space. Let $\{\vp_a^{cl}(\vec y)\}$
denote some
particular classical solution of the equations of motion at $T=0$. We 
assume that for every such classical solution there is a
BCFT describing open string propagation in this background
$\{\vp_a^{cl}\}$. In that case,
we can formulate string field
theory around this new background and define a tachyonic mode around this
background using the prescription of the last section. We shall choose the
coordinate $T$ appearing in $\LL(\{\vp_a(\vec y)\},T)$ in such a way that
around every classical solution, keeping
$\{\vp_a(\vec y)\}$ fixed at $\{\vp_a^{cl}(\vec y)\}$ and changing $T$
corresponds to switching on the
tachyonic mode(s) of the string field theory formulated in the background
$\{\vp_a^{cl}(\vec y)\}$. This has been schematically illustrated in
Fig.\ref{f1}.
In principle there could be obstruction to such a choice of coordinates;
we shall assume that there is no such obstruction.

Since $\vp_a(\vec y)=0$ denotes a trivial classical solution representing
the original D-brane,
we have, according to eq.\refb{e2.1}
\be \label{e4.1}
\LL(\vp_a=0, T)= - M_0f(T)\, ,
\ee
where $M_0$ denotes the mass of the brane for $\vp_a=0$. We have chosen
the
additive constant in $\LL$ such that $\LL$ vanishes at $\vp_a(\vec y)=0$,
$T=0$.\footnote{This is natural from the point of view of string field
theory formulated in the background BCFT corresponding to $\vp_a=0, T=0$.
On the other hand, from the point of view of the effective action, it is
often more natural to choose
this additive constant in such a way that $\LL(\vp_a=0,T=0)$ is equal to
$-M_0$, $-$ the negative of the mass of the original D-brane. This is what
is done, for example, in writing the
action in the Born-Infeld form.} Let $\vp_a^{cl}$ denote a non-trivial
classical solution of the equations of motion representing a new BCFT, and
$M$ denote the mass of the D-brane described by this new BCFT.\footnote{If
we consider a new BCFT with the open string coupling constant fixed, then
the mass of the D-brane does not depend on the BCFT. But it is more
natural to keep the closed string coupling constant (dilaton) fixed as we
change the open string background. Since the relationship between the
closed and the open string coupling constant does depend on the
BCFT\cite{9908142}, the D-brane in the new background can have a different
mass.}
According to the result
of the previous section, the effective lagrangian of $T$, formulated
around the new background, should be given by $-Mf(T)$. This gives,
\be \label{e4.2} 
\LL(\vp_a^{cl}, T)= -Mf(T) + K\, , 
\ee
where $K$ is an additive constant. The origin of this constant may be
understood as follows. In defining effective lagrangian $\LL$, we have
fixed the
additive constant in the action in such a way that the lagrangian vanishes
when all the fields are set to zero. In this convention, if
$\{\vp_a^{cl}\}$ denotes a
time independent classical solution of the equations of motion reflecting
a
new BCFT, then the value of the lagrangian of the original string field
theory, evaluated at $\vp_a=\vp_a^{cl}$, will reflect the difference
between the potential energies of the initial and the final
configurations. On the other hand the effective lagrangian obtained
by integrating out the degress of freedom of the string field
theory action formulated
directly around the new BCFT will have zero value when all the fields in
this {\it new string field theory action} are set to zero. Thus the two
effective lagrangians
must differ by an additive constant $K$. It is fixed
by demanding
that 
\be \label{eldiff} 
\LL(\vp_a^{cl},T=0)-\LL(\vp_a=0,T=0)=-(M-M_0)\, .
\ee
Since $f(0)=0$, this gives, using eqs.\refb{e4.1} and \refb{e4.2}
\be \label{ekeq}
K= M_0-M\, .
\ee
Hence
\be \label{eleq}
\LL(\vp_a^{cl}, T)= -M(1+f(T)) + M_0\, .
\ee
Using
eqs.\refb{eleq} and \refb{efvan} we see that, 
\be \label{e4.3}
\LL(\vp_a^{cl}, T_0)=M_0\, .
\ee
In other words the lagrangian at $T=T_0$ has the same value $M_0$ for all
$\{\vp_a^{cl}(\vec y)\}$ which correspond to solutions of the equations of
motion at
$T=0$. 
Although this does not prove that $\LL(\vp_a,T_0)$ is independent of
$\vp_a$
(and hence in particular of the U(1) gauge fields) for all $\vp_a$, it
certainly lends support to this conjecture. 

In the specific context of the
U(1) gauge field, note that
if $F_{mn}$ denote the components of the U(1) gauge field strength on the
D-brane,
then since constant $F_{mn}$ is a solution of the equations of motion and
describes a BCFT, the lagrangian at $T=T_0$ is independent of $F_{mn}$
at least for
constant $F_{mn}$.  Thus at $T=T_0$, $\LL$ can at most contain terms
involving derivatives of $F_{mn}$.
This establishes that $\LL(F_{mn},T_0)$ does
not contain the standard gauge kinetic term since it vanishes for constant
$F_{mn}$, and hence even if $\LL$ is not completely independent of 
$F_{mn}$ at $T=T_0$, it does not represent
a standard gauge theory.

The fact that $\LL(F_{mn},T_0)$ does not depend on $F_{mn}$ for constant
$F_{mn}$ can also be seen via a T-duality transformation, starting with
the assumption that at $F_{mn}=0$ the mass of the brane, $-\LL+M_0$,
vanishes at the extremum $T_0$ of the tachyon potential. For
this let $x^1$ and $x^2$ denote two of the directions tangential to the
D-brane
which have been compactified. For $F_{mn}=0$, an $R\to(1/R)$ duality
transformation along the $x^2$ direction converts this D-brane to a
D-brane
with Dirichlet boundary condition along the $x^2$ direction, and Neumann
boundary condition along the $x^1$ direction. Since the mass of the 
brane does not change under T-duality, the mass 
of the T-dualized brane, and hence also its tension,
vanishes at the extremum $T_0$ of the tachyon potential. Now if we
switch on the constant field strength
$F_{12}$ in the original D-brane, it corresponds to putting Dirichlet
boundary condition on some linear combination of $x^1$ and $x^2$ in the
T-dual description. Thus we effectively change the orientation of the
brane in the T-dual description. But if the tension of this D-brane
vanishes at some
extremum of the tachyon potential, it continues to vanish even if we
change
the orientation of the brane, and hence the total mass of the brane still
vanishes. But this is equal to the mass of the original brane at constant
$F_{12}$ and $T=T_0$, {\it i.e.} to $-\LL(F_{12},T_0)+M_0$. Thus we see
that
$\LL(F_{12},T_0)=M_0$, {\it i.e.} it is independent of $F_{12}$.

This analysis can be easily generalized to the case of the brane-antibrane
system
and the non-BPS D-brane of type II string theories. In carrying out this
analysis one should keep in mind that for the brane-antibrane system, the
U(1) which must be switched on is the diagonal combination of the two
U(1)'s on the brane and the antibrane (corresponding to CP sector $I$) so
that the new BCFT satisfies the conditions for validity of our analysis.
It is only for this U(1) that we conjecture that the
action is independent of the gauge field at
the minimum of the tachyon potential. The other U(1) gauge symmetry is
broken due to Higgs mechanism in the presence of a non-vanishing vev of
the tachyon field.

\sectiono{Tachyon potential in closed bosonic string theory} \label{s5}

We can repeat our analysis for the tachyon of closed bosonic string theory
in arbitrary conformal field theory background. In this case a string
field configuration is represented by an arbitrary state $|\Phi\rrr$ in
the closed
string Hilbert space carrying ghost number 2, and satisfying the condition
\be \label{e5.1}
(b_0-\bar b_0)|\Phi\rrr = 0, \qquad (L_0-\bar L_0)|\Phi\rrr=0 \, .
\ee
There is an action similar to \refb{e3.1} for the closed string field
theory, with the difference that the action is
non-polynomial\cite{SAADI,9305026}, involving quartic and higher order
vertices. However, each of these vertices are constructed from conformal
field theory correlation functions in a manner analogous to \refb{e3.1}.
Thus
we can find a consistent truncation of the theory by restricting the
string field configuration to a subspace $\HH_1$ built from $|0\rrr$ by
the action of the ghost oscillators and the matter Virasoro generators.

The
zero momentum tachyon corresponds to the state $c_1\bar c_1|0\rrr$, and
hence is an element of $\HH_1$. Thus starting from the truncated action
and integrating out the other fields we can recover the tachyon
potential.\footnote{In this case the zero momentum massless dilaton,
corresponding to the state $(c_{-1}c_1-\bar c_{-1}\bar c_1)|0\rrr$ also
belongs to the set $\HH_1$ and cannot be integrated out. Thus by this
procedure we shall get the potential involving the tachyon and the zero
momentum dilaton.}
This is insensitive to the details of the conformal field theory on which
the bosonic string theory is based, and thus is universal. However, unlike
in the
case of open string tachyons, in this case there is no compelling reason
to believe that there exists a non-trivial classical solution of the
string field theory equations of motion in this truncated theory; hence
the physical significance of the tachyon potential obtained this way is
not entirely obvious.

{\bf Acknowledgement}: I wish to thank A.~Dabholkar and B.~Zwiebach for
useful
discussions.

\end{document}